\documentclass[%
 reprint,
 amsmath,amssymb,
 aip,
]{revtex4-1}

\usepackage{graphicx}
\usepackage{upgreek}
\usepackage{dcolumn}
\usepackage{bm}
\usepackage{verbatim}
\usepackage{textcomp}
\usepackage{hyperref}
\usepackage{natbib}

\begin{document}

\preprint{APS/123-QED}

\title{Influence of elastically pinned magnetic domain walls on magnetization reversal in multiferroic heterostructures} 

\author{Arianna Casiraghi}
\affiliation{NanoSpin, Department of Applied Physics, Aalto University School of Science, P.O. Box 15100, FI-00076 Aalto, Finland}
\author{Teresa Rinc\'{o}n Dom\'{i}nguez}
\affiliation{NanoSpin, Department of Applied Physics, Aalto University School of Science, P.O. Box 15100, FI-00076 Aalto, Finland}
\author{Stefan R\"{o}\ss ler}
\affiliation{Universit\"{a}t Hamburg, Institut f\"{u}r Angewandte Physik, Jungiusstr. 11, 20355 Hamburg, Germany}
\author{K\'{e}vin J. A. Franke}
\affiliation{NanoSpin, Department of Applied Physics, Aalto University School of Science, P.O. Box 15100, FI-00076 Aalto, Finland}
\author{Diego L\'{o}pez Gonz\'{a}lez}
\affiliation{NanoSpin, Department of Applied Physics, Aalto University School of Science, P.O. Box 15100, FI-00076 Aalto, Finland}
\author{Sampo J. H\"{a}m\"{a}l\"{a}inen}
\affiliation{NanoSpin, Department of Applied Physics, Aalto University School of Science, P.O. Box 15100, FI-00076 Aalto, Finland}
\author{Robert Fr\"{o}mter}
\affiliation{Universit\"{a}t Hamburg, Institut f\"{u}r Angewandte Physik, Jungiusstr. 11, 20355 Hamburg, Germany}
\author{Hans Peter Oepen}
\affiliation{Universit\"{a}t Hamburg, Institut f\"{u}r Angewandte Physik, Jungiusstr. 11, 20355 Hamburg, Germany}
\author{Sebastiaan van Dijken}
\email{sebastiaan.van.dijken@aalto.fi}
\affiliation{NanoSpin, Department of Applied Physics, Aalto University School of Science, P.O. Box 15100, FI-00076 Aalto, Finland}

\begin{abstract}
In elastically coupled multiferroic heterostructures that exhibit full domain correlations between ferroelectric and ferromagnetic sub-systems, magnetic domain walls are firmly pinned on top of ferroelectric domain boundaries. In this work we investigate the influence of pinned magnetic domain walls on the magnetization reversal process in a Co$_{40}$Fe$_{40}$B$_{20}$ wedge film that is coupled to a ferroelectric BaTiO$_{3}$ substrate via interface strain transfer. We show that the magnetic field direction can be used to select between two distinct magnetization reversal mechanisms, namely (1) double switching events involving alternate stripe domains at a time or (2) synchronized switching of all domains. Furthermore, scaling of the switching fields with domain width and film thickness is also found to depend on field orientation. These results are explained by considering the dissimilar energies of the two types of pinned magnetic domain walls that are formed in the system.

\end{abstract}

\maketitle

\section{\label{sec:level1}Introduction}
The magnetic properties of a ferromagnetic film are determined by the relationship between intrinsic material parameters, such as exchange stiffness and magnetocrystalline anisotropy, and extrinsic effects, including magnetic anisotropies induced by shape, strain and interfaces. Typically, both intrinsic and extrinsic contributions give rise to a magnetic energy landscape that is uniform across the ferromagnetic film, aside from local variations caused by defects and/or film roughness. The magnetization reversal process and the magnetic hysteresis loop do not therefore usually depend on probing area. In conventional magnetic films with uniform magnetic anisotropy, magnetic switching proceeds by nucleation of reversed domains and subsequent domain growth via lateral domain wall motion. Models describing thermally activated magnetization reversal and magnetic domain wall motion use parameters that vary with magnetic anisotropy energy and random fluctuations thereof, while the energetics of magnetic domain walls is not specifically taken into account \cite{Labrune, Bruno, Lemerle}. Since the density of domain walls is often low and their spin structure and energy remain nearly constant during domain growth, this omission is justified for most ferromagnetic systems. However, if the motion of magnetic domain walls is prohibited by strong pinning, the energetics of domain walls can have a more pronounced influence on magnetization reversal, especially when the density of pinned walls is high and the anisotropy axes in neighbouring domains are non-collinear.

Strong local pinning of magnetic domain walls can be attained by various methods, including focused ion beam or low-energy proton irradiation \cite{Chappert, Terris, Fassbender2004, Fassbender2008, Franken, Kim, Hamann} and oxygen ion migration from an adjacent metal-oxide layer \cite{Bauer2013, Bauer2015}. Other promising strategies to locally tailor the magnetic properties of a continuous magnetic medium exploit exchange coupling with a multiferroic BiFeO$_{3}$ layer \cite{Chu, Lebeugle, Heron, You} or strain coupling to the ferroelastic domains of a ferroelectric BaTiO$_{3}$ substrate \cite{Lahtinen2011, Lahtinen2012, Franke2012, Chopdekar, Streubel, Franke2014}. In both cases, one-to-one correlations between the domains in BiFeO$_{3}$ or BaTiO$_{3}$ and the domains of an adjacent ferromagnetic film have been demonstrated. Furthermore, since the magnetic domain walls are firmly pinned on top of ferroelectric domain boundaries by abrupt changes in magnetic anisotropy, they do not move during magnetization reversal \cite{Lahtinen2011, Lahtinen2012}. This strong pinning effect leads to the formation of two types of magnetic domain walls with considerably different energy, depending on the direction of in-plane magnetic field \cite{Franke2012, Franke2014}.

In this work we investigate the influence of pinned magnetic domain walls on magnetization reversal in a strain-coupled Co$_{40}$Fe$_{40}$B$_{20}$/BaTiO$_{3}$ heterostructure with regular magnetic stripe domains (Fig. \ref{Fig1}). We find that magnetic switching in this system depends strongly on the type of magnetic domain wall that is created during a magnetic field sweep, especially if the thickness of the CoFeB film exceeds 50 nm. For magnetic fields along the stripe domains, high-energy head-to-head and tail-to-tail domain walls form. In this case, magnetization reversal proceeds in two clear steps involving abrupt magnetic switching in every second stripe domain at a time. The regular lateral modulations in the magnetization reversal process are driven by transformations of the domain wall structure into a low-energy head-to-tail configuration. On the other hand, if the magnetic field is applied perpendicular to the stripe domains, low-energy head-to-tail domain walls form. As a consequence, domain wall transformations cannot reduce the energy of the system and the magnetization of each domain switches simultaneously. The dependence of the magnetic switching fields for both reversal mechanisms on CoFeB film thickness and stripe domain width are discussed in detail.

\section{\label{sec:level2}Experimental details}

The experiments were conducted on a multiferroic heterostructure composed of a ferroelectric BaTiO$_{3}$ (001) single-crystal substrate and a ferromagnetic Co$_{40}$Fe$_{40}$B$_{20}$ wedge film, with thickness $t$ = 0 -- 150 nm. The wedge film was deposited via magnetron sputtering at 300 $^{\circ}$C. Upon cooling through the paraelectric-to-ferroelectric phase transition at 120 $^{\circ}$C, the lattice structure of BaTiO$_{3}$ becomes tetragonal and regular a$_{1}$ -- a$_{2}$ ferroelastic stripe domains are formed to minimize electrostatic and elastic energies \cite{Merz}. The alternating 90$^{\circ}$ in-plane rotations of the lattice tetragonality that are characteristic of this domain pattern give rise, via inverse magnetostriction, to 
corresponding 90$^{\circ}$ in-plane rotations of the uniaxial magnetoelastic anisotropy axis in the CoFeB film.

\begin{figure}[!ht]
\includegraphics[width=7cm]{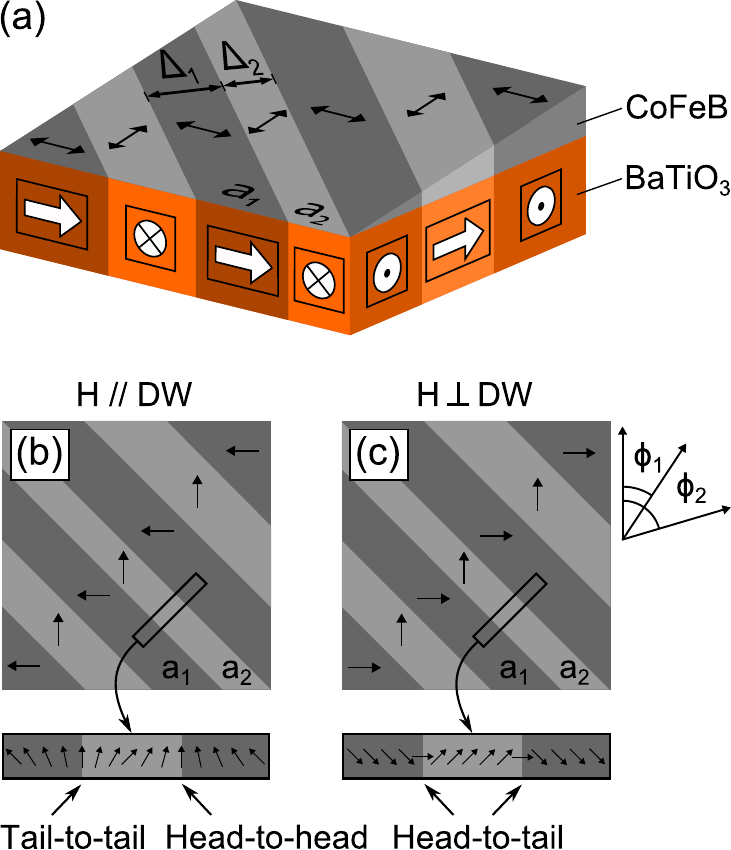}
\caption{(a) Sketch of the multiferroic heterostructure: white arrows indicate the ferroelectric polarization direction of a$_{1}$--a$_{2}$ domains in BaTiO$_{3}$, while black arrows indicate the direction of the uniaxial magnetoelastic anisotropy axes of the corresponding a$_{1}$--a$_{2}$ domains in CoFeB. The widths $\Delta_{1}$ and $\Delta_{2}$ of a$_{1}$ and a$_{2}$ domains range between 1 \textmu m and 10 \textmu m (on average $\Delta_{1}$ $>$ $\Delta_{2}$). Sketch of the configuration of 90$^{\circ}$ magnetically charged (b) and uncharged (c) domain walls obtained at remanence, after reducing the magnetic field from saturation along the direction parallel and perpendicular to the domain walls, respectively. ${\phi_{1}}$ and $\phi_{2}$ represent the magnetization angles in the a$_{1}$ and a$_{2}$ domains.}
\label{Fig1}
\end{figure}

At room temperature the ferroelastic a$_{1}$ and a$_{2}$ stripe domains are found to be fully imprinted into the CoFeB wedge film at all thicknesses, as schematically illustrated in Fig. \ref{Fig1}(a). This indicates that the magnetoelastic anisotropy dominates over the other anisotropy contributions in CoFeB even at the thick side of the film. 
The ferromagnetic domains were imaged at room temperature using magneto-optical Kerr effect (MOKE) microscopy with in-plane magnetic field. Magnetic hysteresis loops of individual a$_{1}$ and a$_{2}$ stripe domains were extracted from the variation of local magnetic contrast during magnetization reversal.

\section{\label{sec:level3}Magnetically charged and uncharged domain walls}

The magnetic domain walls that separate the imprinted a$_{1}$ and a$_{2}$ domains in CoFeB are strongly pinned onto the ferroelectric domain boundaries by the sudden rotation of the uniaxial magnetic anisotropy axes. As a result of this pinning effect, the magnetic domain walls do not move under application of a magnetic field and the total spin rotation within the walls varies with the direction and strength of the in-plane magnetic field \cite{Franke2012}. 
Particularly relevant to the present work is the possibility to initialize magnetic domain walls with two distinct spin structures for specific magnetic field orientations. When the field is reduced from saturation along the direction parallel to the walls,  the spins align in alternating head-to-head and tail-to-tail configurations which induce magnetostatic charges on each side of the domain walls (Fig. \ref{Fig1}(b)). Accordingly, these domain walls are usually referred to as magnetically charged. On the other hand, when the field is reduced from saturation along the direction perpendicular to the walls, magnetically uncharged head-to-tail domain walls are instead formed (Fig. \ref{Fig1}(c)). In ferromagnetic films without anisotropy modulations charged walls typically arrange in zigzag configurations to reduce magnetostatic charge density \cite{Hubert1979, Hubert1998, Favieres}. In our strain-coupled multiferroic heterostructure, however, both magnetically charged and uncharged walls are perfectly straight because of strong pinning onto the underlying ferroelectric domain boundaries.  

The profile of charged magnetic domain walls is mostly determined by the competition between magnetostatic energy and magnetic anisotropy energy, while exchange energy and magnetic anisotropy energy mainly define the structure of uncharged magnetic domain walls \cite{Franke2014}. As magnetostatic coupling between spins extends over a longer distance than exchange interactions, the width and energy of charged walls are considerably larger than that of uncharged walls. Moreover, since the magnetostatic energy increases with ferromagnetic film thickness, the difference between the width and energy of charged and uncharged domain walls becomes more pronounced for thick films\cite{Franke2014}.  

\begin{figure}[!ht]
\includegraphics[]{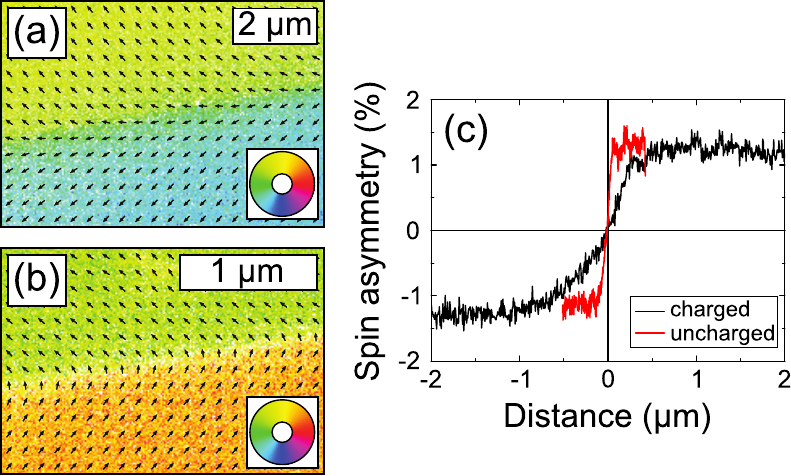} 
\caption{Two-component SEMPA images of the remanent in-plane magnetization configuration of (a) charged and (b) uncharged magnetic domain walls in a 20 nm CoFeB/BaTiO$_{3}$ heterostructure. (c) Comparison of the domain wall profiles as obtained from a line scan along the direction perpendicular to the walls in (a) and (b). The spin asymmetry is proportional to the projection of the magnetization along the scan direction, for the charged wall, and perpendicular to it, for the uncharged wall.}
\label{Fig2}
\end{figure}

Fig. \ref{Fig2}(a) and (b) show Scanning Electron Microscopy with Polarization Analysis (SEMPA) images of charged and uncharged magnetic domains walls, at remanence, in a 20 nm CoFeB film that was grown on BaTiO$_{3}$ under the same conditions as the wedge film. 
The elevated spatial resolution of SEMPA \cite{Berger, Fromter} allows for domain wall imaging at the nanoscale and for the extraction of the corresponding wall profiles (Fig. \ref{Fig2}(c)). Following the domain wall width definition of Lilley \cite{Lilley}, the widths of the charged and uncharged domain walls are estimated as $\delta_{c} = 770 \pm 60$ nm and $\delta_{uc} = 165 \pm 25$ nm, respectively.

\section{\label{sec:level3} Results and discussion}

\begin{figure}[!ht]
\includegraphics[width=7.5cm]{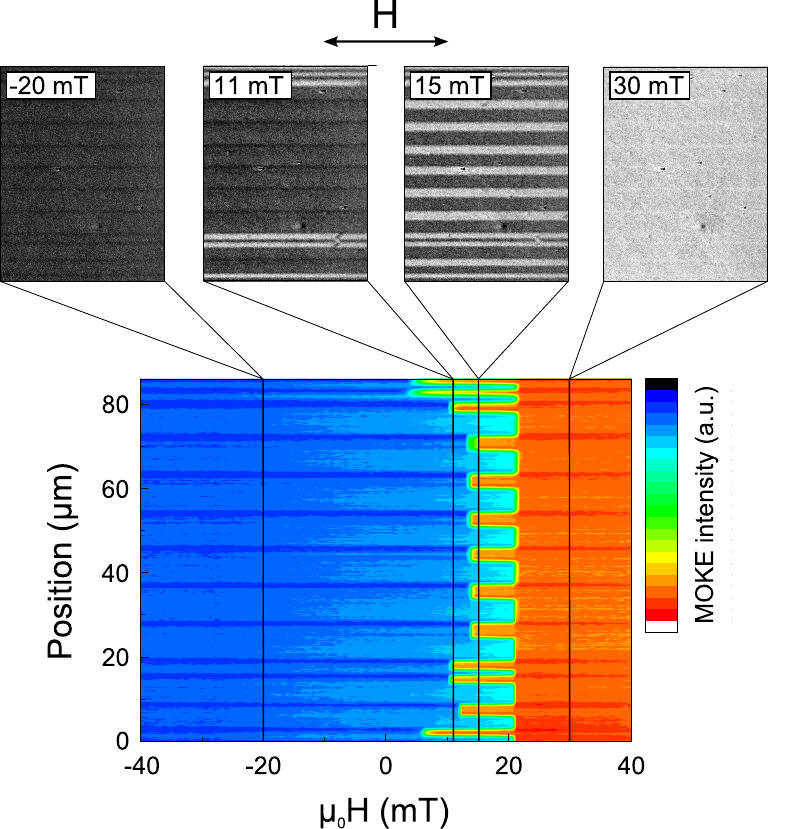}
\caption{Visualization of the averaging process used to combine a set of standard MOKE images, measured while sweeping the field from negative to positive saturating values (in this case along the domain walls), into a single contour plot displaying MOKE intensity as a function of position and magnetic field strength. Blue (red) color corresponds to magnetization pointing to the left (right).}
\label{Fig3}
\end{figure}

The magnetization reversal process in CoFeB under the formation of charged or uncharged domain walls is investigated by MOKE microscopy.
To this end, a large set of MOKE images is collected as a function of magnetic field strength. This is done with  the magnetic field applied either parallel or perpendicular to the domain walls and for different CoFeB film thicknesses. For each field sweep, the MOKE intensity of individual images is averaged along the direction of the stripe domains and combined into a single contour plot, as illustrated in Fig. \ref{Fig3}. 
Such contour plots, wherein a vertical line contains information about the magnetization direction of each MOKE image, are used here as an efficient way to visualize magnetic switching in all the domains of the original set of images. 

\begin{figure*}[!ht]
\includegraphics[width=17.8cm]{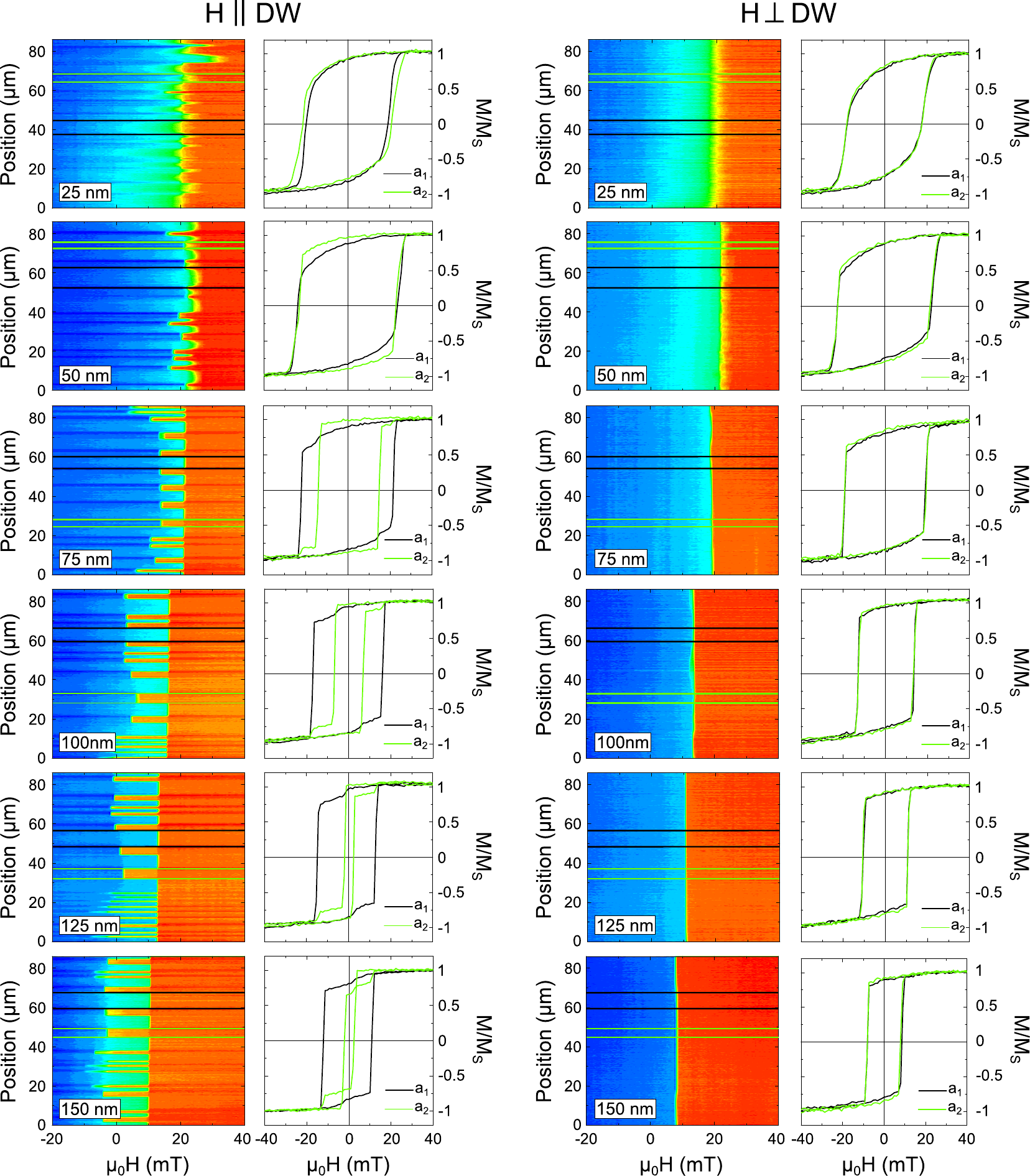}
\caption{Contour plots illustrating the magnetization reversal process at several film thicknesses while sweeping the magnetic field $H$ from negative to positive saturating values along the direction either parallel (left column) or perpendicular (right column) to the domain walls (DWs). The sample areas imaged in the left and right columns are the same. The scale of the MOKE intensity is identical to the one in Fig. \ref{Fig3}. When $H \parallel$ DWs blue (red) color corresponds to magnetization pointing to the left (right), while when $H \perp$ DWs blue (red) color corresponds to magnetization pointing down (up). The black and green solid lines in the contour plots mark the boundaries of the a$_{1}$ and a$_{2}$ domains, respectively, whose magnetic hysteresis loop is shown to the side.}
\label{Fig4}
\end{figure*}

MOKE contour plots with magnetic field applied either parallel or perpendicular to the domain walls are shown in Fig. \ref{Fig4} for CoFeB film thicknesses ranging between 25 nm and 150 nm.  
To facilitate direct comparisons, the magnetic hysteresis loops of selected a$_{1}$ and a$_{2}$ domains are also shown. The width of the selected a$_{1}$ and a$_{2}$ domains is similar for all thicknesses, being on average $\Delta_{1} \sim 8$ \textmu m and $\Delta_{2} \sim 5$ \textmu m, respectively.
In both cases, the magnetic field is applied at an angle of 45$^{\circ}$ with respect to the easy axes of a$_{1}$ and a$_{2}$ domains (see Fig. \ref{Fig1}). The reversal process should therefore not depend on the field direction nor the domain type, in accordance with the Stoner--Wohlfarth model for single domains with uniaxial magnetic anisotropy. In our system, however, the magnetic switching behavior depends strongly on the direction of the magnetic field and, for field parallel to the walls, on the domain type. 
In particular, a$_{1}$ and a$_{2}$ domains switch simultaneously when the magnetic field is applied perpendicularly to the walls, while they switch at distinct fields when the magnetic field is applied parallel to the walls. This unusual switching behavior is caused by strong magnetic domain wall pinning and the substantial energy difference between charged and uncharged walls \cite{Hubert1979, Franke2014}, as discussed below. 

\begin{figure*}[!ht]
\includegraphics[width=16 cm]{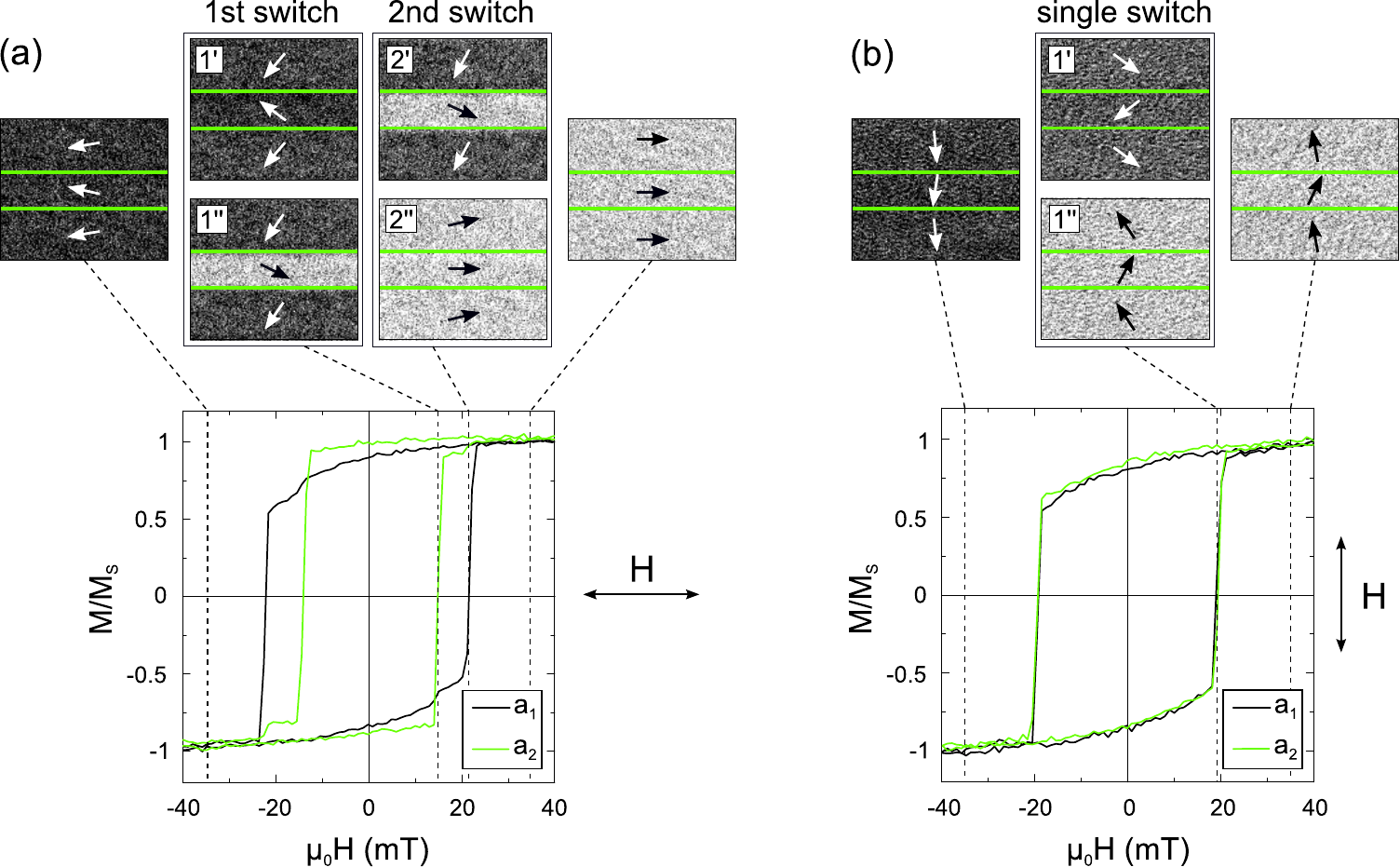}
\caption{MOKE images illustrating the reversal mechanism of a portion of the film  (comprising one a$_{2}$ domain surrounded by two a$_{1}$ domains) at a thickness $t$ = 75 nm, while sweeping the field from negative to positive saturating values along the direction parallel (a) and perpendicular (b) to the domain walls. The arrows in the images indicate the direction of the magnetization of each domain, as derived from the MOKE intensity. The green solid lines mark the boundaries of the a$_{2}$ domain. Each image is linked via a dashed line to the corresponding position along the hysteresis loops.}
\label{Fig5}
\end{figure*}

The influence of the two types of domain walls on the reversal process is clarified through consideration of individual MOKE images illustrating the switching of a$_{1}$ and a$_{2}$ domains at a specific thickness ($t$ = 75 nm), as shown in Fig. \ref{Fig5}. 
When the magnetic field is applied parallel to the walls (Fig. \ref{Fig5}(a)) and reduced from saturation, the magnetization of each stripe domain rotates towards the respective easy anisotropy axes, causing the formation of charged domain walls. Both the energy and width of these walls are initially small, but rapidly increase upon decreasing field strength. At some field value (Fig. \ref{Fig5}(a--$1'$)), the energy of charged walls becomes so large that an abrupt magnetic switching event is triggered in every second stripe domain (Fig. \ref{Fig5}(a--$1''$)), here defined as a$_{2}$. During this first switching event all charged domain walls transform into lower-energy uncharged domain walls, thus providing a net energy gain for the entire magnetic system. Upon a further increase of the magnetic field strength, the a$_{1}$ domains switch too: during this second switching event all uncharged walls (Fig. \ref{Fig5}(a--$2'$) are transformed back into charged walls (Fig. \ref{Fig5}(a--$2''$)) which are now characterized by a modest spin rotation and, thus, considerably smaller energy. A different reversal process occurs when the magnetic field is applied perpendicular to the domain walls (Fig. \ref{Fig5}(b)): now, uncharged walls are formed when the magnetic field is reduced from saturation and the magnetic system cannot reduce its energy by domain wall transformations. Instead, in order to prevent the formation of higher-energy charged walls, magnetic switching is now completely synchronized in all domains (Fig. \ref{Fig5}(a--$1'$) and (a--$1''$)). A much weaker dependence of the magnetic hysteresis curve on the direction of applied magnetic field has been reported for an exchange-coupled La$_{0.7}$Sr$_{0.3}$MnO$_{3}$/BiFeO$_{3}$ heterostructure \cite{You}.

\subsection{Scaling of magnetic switching with domain width }

\begin{figure}[!ht]
\includegraphics[width=8.5cm]{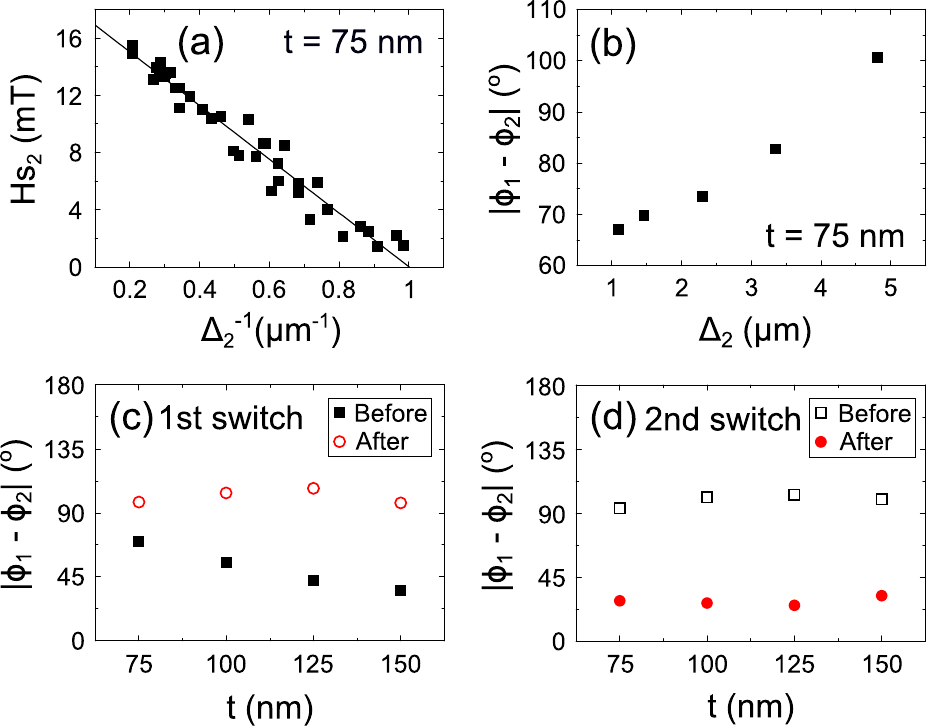}
\caption{(a) Dependence of the switching field of the a$_{2}$ domains on domain width, measured at $t$ = 75 nm. Symbols represent experimental data while the line is the corresponding linear fit. (b) Dependence of the spin rotation of the charged domain walls that form before the reversal of the a$_{2}$ domains (first switching event) as a function of domains width, again for $t$ = 75 nm. Dependence of the spin rotation of the domain walls that form before and after the first (c) and second (d) switching event as a function of thickness, for $\Delta_{2} \sim 5$ \textmu m. Closed (open) symbols in (c) and (d) indicate magnetically charged (uncharged) domain walls.
 }
\label{Fig6}
\end{figure}

An intriguing aspect that emerges from the contour plots in Fig. \ref{Fig4} is the influence of domain width on the magnetization reversal process. When the magnetic field is applied along the stripe domains, charged walls are created and the switching field of the a$_{2}$ domains ($H_{S2}$) decreases with decreasing width $\Delta_{2}$. Specifically, $H_{S2}$ is inversely proportional to $\Delta_{2}$, as illustrated in Fig. \ref{Fig6}(a) for $t$ = 75 nm. 
This scaling behavior is explained considering that narrow a$_{2}$ domains correspond to a higher density of charged domain walls and, consequently, a higher magnetic energy density compared to wide a$_{2}$ domains. 
An alternative way to understand this behavior is provided in Fig. \ref{Fig6}(b), where the value of the spin rotation of charged walls, measured just before the first switching event, is plotted as a function of $\Delta_{2}$, for $t$ = 75 nm.  For $\Delta_{2}$ = 5 \textmu m  the spin rotation of charged domain walls increases up to $100^{\circ}$ before switching, while in narrower a$_{2}$ domains charged walls with considerably smaller spin rotation already transform into uncharged walls. Because larger spin rotations are associated with higher magnetic energy, Fig. \ref{Fig6}(b) indicates that wider a$_{2}$ domains can accommodate charged domain walls with higher energy than narrow  a$_{2}$ domains, before transformations to uncharged domain walls occur.  

While $H_{S2}$ scales with $\Delta_{2}$ when the field is applied along the stripe domains, $H_{S1}$ is independent of $\Delta_{1}$. Considering the arguments that were provided to explain the dependence of $H_{S2}$ on $\Delta_{2}$, this circumstance may appear in contradiction with the fact that uncharged walls are transformed back into charged walls when a$_{1}$ domains switch (see the second switching event in Fig. \ref{Fig5} (a)). However, charged walls that form after the second switching event have much smaller spin rotations, and correspondingly lower magnetic energies, than the charged walls before the first switching event. This is illustrated in Figs. \ref {Fig6} (c) -- (d), where the spin rotation of charged and uncharged walls before and after the two switching events is plotted as a function of CoFeB film thickness, for a$_{2}$ domains with similar width $\Delta_{2} \sim 5$ \textmu m (a$_{1}$ and a$_{2}$ domains switch almost simultaneously for $t \leq$ 50 nm and the corresponding data have been omitted from these figures). Since the energy of charged walls is small after magnetic switching in the a$_{1}$ domains, $H_{S1}$ is mainly determined by the magnetic anisotropy inside the domain, which is independent of $\Delta_{1}$, rather than the energetics of the domain walls. This observation is also confirmed by the fact that $H_{S1}$ is nearly identical for the two field orientations (see Fig. \ref{Fig4}).

\subsection{Scaling of magnetic switching with CoFeB film thickness}

\begin{figure}
\includegraphics[width=7.5cm]{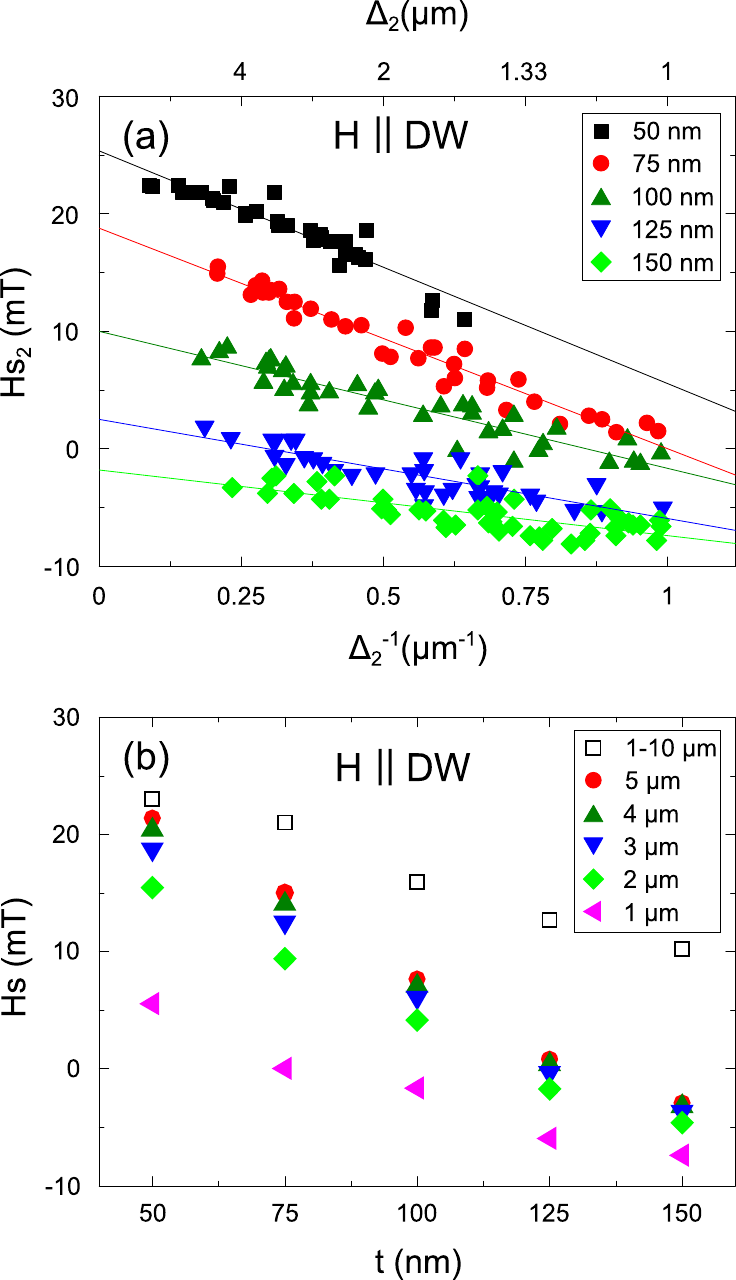}
\caption{(a) Dependence of the switching field of the a$_{2}$ domains on domain width and CoFeB film thickness, for thicknesses $\geq$ 50 nm. Symbols represent experimental data while lines are the corresponding linear fits. (b) Comparison between the switching fields of a$_{1}$ domains (open symbols) and a$_{2}$ domains (closed symbols) as a function of film thickness and for $\Delta_{2}$ values ranging between 1 \textmu m and 5 \textmu m (the switching field is independent of $\Delta$ in the a$_{1}$ domains). In both (a) and (b) the magnetic field is applied parallel to the domain walls.}
\label{Fig7}
\end{figure}

Finally, the dependence of the magnetic switching field on CoFeB film thickness is discussed. From Fig. \ref{Fig4} it can be seen that both $H_{S1}$ and $H_{S2}$ decrease with CoFeB thickness for t $\geq$ 50 nm. The switching fields of a$_{1}$ domains (irrespective of field direction) and a$_{2}$ domains (magnetic field perpendicular to the walls) are not drastically influenced by domain wall transformations and thus their variation with film thickness mimics that of ferromagnetic films without regular anisotropy modulations. A different dependence of $H_{S2}$ on film thickness is obtained when charged magnetic walls are formed. In this case, two additional interlinked parameters affect $H_{S2}$, namely (1) the width of the a$_{2}$ domains (Fig. \ref{Fig6} (a)) and (2) the spin rotation within the walls and thus the wall energy (Figs. \ref {Fig6} (b) -- (d)). The dependence of $H_{S2}$ on film thickness and $\Delta_{2}$ is summarized in Figs. \ref{Fig7} (a) and (b). 
For relatively thin films, the energy difference between charged and uncharged magnetic domain walls is rather modest, giving rise to large spin rotations inside both domain walls and strong scaling of $H_{S2}$ with $1/\Delta_{2}$. At the other end of the thickness spectrum, the width and energy of uncharged walls are mostly unchanged, whereas those of charged walls are significantly enhanced. The growing energy difference between the two types of domain walls leads to a reduction of $H_{S2}$ with increasing CoFeB thickness. This scaling effect is most significant for wide a$_{2}$ domains, since charged walls can attain their full width and energy without restrictions. For small $\Delta_{2}$, however, the domain walls are artificially confined to narrow domain stripes, leading to a finite-size reduction of the spin rotation within the walls and thus of their width and energy \cite{Franke2014}. This effect hampers the reduction of $H_{S2}$ with film thickness in narrow domains and, consequently,  the scaling of $H_{S2}$ with $1/\Delta_{2}$ becomes less pronounced at the thick side of the CoFeB wedge. When the energy gain during charged-to-uncharged wall transformations exceeds the magnetic anisotropy energy of the domains, magnetic switching in the a$_{2}$ domains can take place before zero applied magnetic field is reached. Negative switching fields are measured for narrow a$_{2}$ domains and/or at large CoFeB film thicknesses, as shown in Figs. \ref{Fig7} (a) and (b). 

The $1/\Delta_{2}$ scaling behavior in our multiferroic heterostructure is qualitatively similar to that of two-dimensional magnetic systems with opposing interface and bulk anisotropies. In magnetic multilayers, for example, the interface anisotropy dominates the total energy of the system for very thin ferromagnetic films, but its contribution decays with thickness ($t$) as $1/t$ (see ref. \cite{Johnson}). This scaling effect causes a spin reorientation transition at a critical thickness $t_{c}$. Exchange bias in ferromagnetic-antiferromagnetic bilayers is another well-known magnetic interface effect \cite{Nogues}. The switching field in this case can also be negative (for one field sweep direction) and the magnitude of the exchange bias field scales as $1/t$. In our system, the a$_{2}$ domains are bordered by two straight and strongly pinned domain walls and the possible transformation of their internal spin structure acts as an interface-like potential energy. Consequently, the influence of the domain walls on $H_{S2}$ diminishes with domain width as $1/\Delta_{2}$. Since the energy difference between charged and uncharged magnetic domain walls increases with ferromagnetic film thickness, both vertical and lateral scaling effects can be utilized to tailor micromagnetic switching effects in fully correlated multiferroic heterostructures.

\section{Conclusion}

To conclude, we have investigated how magnetic domain wall pinning in a strain-coupled CoFeB/BaTiO$_{3}$ heterostructure affects the magnetization reversal process. Depending on the in-plane field direction, two distinct mechanisms are identified. If magnetically charged domain walls with high energy are formed during a field sweep, the system lowers its energy via domain wall transformation into uncharged configurations, which corresponds to abrupt magnetic switching in every second stripe domain. This lateral modulation of magnetization reversal is unusual and can result in switching before zero applied magnetic field is reached. On the other hand, rotation of the applied magnetic field by 90$^{\circ}$ results in the formation of low-energy uncharged magnetic domain walls and a very different reversal behavior. In this configuration, all domains switch simultaneously to prevent the formation of high-energy charged magnetic domain walls. The magnetic switching event that is driven by domain wall transformations scales with the energy difference between the two types of magnetic walls and, thus, with the thickness of the CoFeB film. Also, since the local energy of the system varies with the density of magnetic domain walls, the switching field that is associated with domain wall transformations is inversely proportional to the domain width. The observed dependence of magnetization reversal on field direction is anticipated to be a general feature of continuous ferromagnetic films with a regular modulation of non-collinear magnetic anisotropy axes and a high density of pinned magnetic domain walls.\\

This work was supported by the Academy of Finland (Grant No. 260361), the European Research Council (ERC-2012-StG 307502- E-CONTROL) and the ``Deutsche Forschungsgemeinschaft'' via ``Sonderforschungsbereich 668''. 

\bibliographystyle{apsrev}
\bibliography{Biblio}

\begin{thebibliography}{31}
\expandafter\ifx\csname natexlab\endcsname\relax\def\natexlab#1{#1}\fi
\expandafter\ifx\csname bibnamefont\endcsname\relax
  \def\bibnamefont#1{#1}\fi
\expandafter\ifx\csname bibfnamefont\endcsname\relax
  \def\bibfnamefont#1{#1}\fi
\expandafter\ifx\csname citenamefont\endcsname\relax
  \def\citenamefont#1{#1}\fi
\expandafter\ifx\csname url\endcsname\relax
  \def\url#1{\texttt{#1}}\fi
\expandafter\ifx\csname urlprefix\endcsname\relax\def\urlprefix{URL }\fi
\providecommand{\bibinfo}[2]{#2}
\providecommand{\eprint}[2][]{\url{#2}}

\bibitem[{\citenamefont{Labrune et~al.}(1989)\citenamefont{Labrune, Andrieu,
  Rio, and Bernstein}}]{Labrune}
\bibinfo{author}{\bibfnamefont{M.}~\bibnamefont{Labrune}},
  \bibinfo{author}{\bibfnamefont{S.}~\bibnamefont{Andrieu}},
  \bibinfo{author}{\bibfnamefont{F.}~\bibnamefont{Rio}}, \bibnamefont{and}
  \bibinfo{author}{\bibfnamefont{P.}~\bibnamefont{Bernstein}},
  \bibinfo{journal}{J. Magn. Magn. Mater.} \textbf{\bibinfo{volume}{80}},
  \bibinfo{pages}{211} (\bibinfo{year}{1989}).

\bibitem[{\citenamefont{Bruno et~al.}(1990)\citenamefont{Bruno, Bayreuther,
  Beauvillain, Chappert, Lugert, Renard, Renard, and Seiden}}]{Bruno}
\bibinfo{author}{\bibfnamefont{P.}~\bibnamefont{Bruno}},
  \bibinfo{author}{\bibfnamefont{G.}~\bibnamefont{Bayreuther}},
  \bibinfo{author}{\bibfnamefont{P.}~\bibnamefont{Beauvillain}},
  \bibinfo{author}{\bibfnamefont{C.}~\bibnamefont{Chappert}},
  \bibinfo{author}{\bibfnamefont{G.}~\bibnamefont{Lugert}},
  \bibinfo{author}{\bibfnamefont{D.}~\bibnamefont{Renard}},
  \bibinfo{author}{\bibfnamefont{J.~P.} \bibnamefont{Renard}},
  \bibnamefont{and} \bibinfo{author}{\bibfnamefont{J.}~\bibnamefont{Seiden}},
  \bibinfo{journal}{J. Appl. Phys.} \textbf{\bibinfo{volume}{68}},
  \bibinfo{pages}{5759} (\bibinfo{year}{1990}).

\bibitem[{\citenamefont{Lemerle et~al.}(1998)\citenamefont{Lemerle, Ferr\'{e},
  Chappert, Mathet, Giamarchi, and Doussal}}]{Lemerle}
\bibinfo{author}{\bibfnamefont{S.}~\bibnamefont{Lemerle}},
  \bibinfo{author}{\bibfnamefont{J.}~\bibnamefont{Ferr\'{e}}},
  \bibinfo{author}{\bibfnamefont{C.}~\bibnamefont{Chappert}},
  \bibinfo{author}{\bibfnamefont{V.}~\bibnamefont{Mathet}},
  \bibinfo{author}{\bibfnamefont{T.}~\bibnamefont{Giamarchi}},
  \bibnamefont{and} \bibinfo{author}{\bibfnamefont{P.~L.}
  \bibnamefont{Doussal}}, \bibinfo{journal}{Phys. Rev. Lett.}
  \textbf{\bibinfo{volume}{80}}, \bibinfo{pages}{849} (\bibinfo{year}{1998}).

\bibitem[{\citenamefont{Chappert et~al.}(1998)\citenamefont{Chappert, Bernas,
  Ferr\'e, Kottler, Jamet, Chen, Cambril, Devolder, Rousseaux, Mathet
  et~al.}}]{Chappert}
\bibinfo{author}{\bibfnamefont{C.}~\bibnamefont{Chappert}},
  \bibinfo{author}{\bibfnamefont{H.}~\bibnamefont{Bernas}},
  \bibinfo{author}{\bibfnamefont{J.}~\bibnamefont{Ferr\'e}},
  \bibinfo{author}{\bibfnamefont{V.}~\bibnamefont{Kottler}},
  \bibinfo{author}{\bibfnamefont{J.~P.} \bibnamefont{Jamet}},
  \bibinfo{author}{\bibfnamefont{Y.}~\bibnamefont{Chen}},
  \bibinfo{author}{\bibfnamefont{E.}~\bibnamefont{Cambril}},
  \bibinfo{author}{\bibfnamefont{T.}~\bibnamefont{Devolder}},
  \bibinfo{author}{\bibfnamefont{F.}~\bibnamefont{Rousseaux}},
  \bibinfo{author}{\bibfnamefont{V.}~\bibnamefont{Mathet}},
  \bibnamefont{et~al.}, \bibinfo{journal}{Science}
  \textbf{\bibinfo{volume}{280}}, \bibinfo{pages}{1919} (\bibinfo{year}{1998}).

\bibitem[{\citenamefont{Terris et~al.}(1999)\citenamefont{Terris, Folks,
  Weller, Baglin, Kellock, Rothuizen, and Vettiger}}]{Terris}
\bibinfo{author}{\bibfnamefont{B.~D.} \bibnamefont{Terris}},
  \bibinfo{author}{\bibfnamefont{L.}~\bibnamefont{Folks}},
  \bibinfo{author}{\bibfnamefont{D.}~\bibnamefont{Weller}},
  \bibinfo{author}{\bibfnamefont{J.~E.~E.} \bibnamefont{Baglin}},
  \bibinfo{author}{\bibfnamefont{A.~J.} \bibnamefont{Kellock}},
  \bibinfo{author}{\bibfnamefont{H.}~\bibnamefont{Rothuizen}},
  \bibnamefont{and} \bibinfo{author}{\bibfnamefont{P.}~\bibnamefont{Vettiger}},
  \bibinfo{journal}{Appl. Phys. Lett.} \textbf{\bibinfo{volume}{75}},
  \bibinfo{pages}{403} (\bibinfo{year}{1999}).

\bibitem[{\citenamefont{Fassbender et~al.}(2004)\citenamefont{Fassbender,
  Ravelosona, and Samson}}]{Fassbender2004}
\bibinfo{author}{\bibfnamefont{J.}~\bibnamefont{Fassbender}},
  \bibinfo{author}{\bibfnamefont{D.}~\bibnamefont{Ravelosona}},
  \bibnamefont{and} \bibinfo{author}{\bibfnamefont{Y.}~\bibnamefont{Samson}},
  \bibinfo{journal}{J. Phys. D: Appl. Phys.} \textbf{\bibinfo{volume}{37}},
  \bibinfo{pages}{R179} (\bibinfo{year}{2004}).

\bibitem[{\citenamefont{Fassbender and McCord}(2008)}]{Fassbender2008}
\bibinfo{author}{\bibfnamefont{J.}~\bibnamefont{Fassbender}} \bibnamefont{and}
  \bibinfo{author}{\bibfnamefont{J.}~\bibnamefont{McCord}},
  \bibinfo{journal}{J. Magn. Magn. Mater.} \textbf{\bibinfo{volume}{320}},
  \bibinfo{pages}{579} (\bibinfo{year}{2008}).

\bibitem[{\citenamefont{Franken et~al.}(2012)\citenamefont{Franken, Swagten,
  and Koopmans}}]{Franken}
\bibinfo{author}{\bibfnamefont{J.~H.} \bibnamefont{Franken}},
  \bibinfo{author}{\bibfnamefont{H.~J.~M.} \bibnamefont{Swagten}},
  \bibnamefont{and} \bibinfo{author}{\bibfnamefont{B.}~\bibnamefont{Koopmans}},
  \bibinfo{journal}{Nat. Nanotechnol.} \textbf{\bibinfo{volume}{7}},
  \bibinfo{pages}{499} (\bibinfo{year}{2012}).

\bibitem[{\citenamefont{Kim et~al.}(2012)\citenamefont{Kim, Lee, Ko, Son, Kim,
  Kang, and Hong}}]{Kim}
\bibinfo{author}{\bibfnamefont{S.}~\bibnamefont{Kim}},
  \bibinfo{author}{\bibfnamefont{S.}~\bibnamefont{Lee}},
  \bibinfo{author}{\bibfnamefont{J.}~\bibnamefont{Ko}},
  \bibinfo{author}{\bibfnamefont{J.}~\bibnamefont{Son}},
  \bibinfo{author}{\bibfnamefont{M.}~\bibnamefont{Kim}},
  \bibinfo{author}{\bibfnamefont{S.}~\bibnamefont{Kang}}, \bibnamefont{and}
  \bibinfo{author}{\bibfnamefont{J.}~\bibnamefont{Hong}},
  \bibinfo{journal}{Nat. Nanotechnol.} \textbf{\bibinfo{volume}{7}},
  \bibinfo{pages}{567} (\bibinfo{year}{2012}).

\bibitem[{\citenamefont{Hamann et~al.}(2014)\citenamefont{Hamann, Mattheis,
  M\"{o}nch, Fassbender, Schultz, and McCord}}]{Hamann}
\bibinfo{author}{\bibfnamefont{C.}~\bibnamefont{Hamann}},
  \bibinfo{author}{\bibfnamefont{R.}~\bibnamefont{Mattheis}},
  \bibinfo{author}{\bibfnamefont{I.}~\bibnamefont{M\"{o}nch}},
  \bibinfo{author}{\bibfnamefont{J.}~\bibnamefont{Fassbender}},
  \bibinfo{author}{\bibfnamefont{L.}~\bibnamefont{Schultz}}, \bibnamefont{and}
  \bibinfo{author}{\bibfnamefont{J.}~\bibnamefont{McCord}},
  \bibinfo{journal}{New J. Phys.} \textbf{\bibinfo{volume}{16}},
  \bibinfo{pages}{023010} (\bibinfo{year}{2014}).

\bibitem[{\citenamefont{Bauer et~al.}(2013)\citenamefont{Bauer, Emori, and
  Beach}}]{Bauer2013}
\bibinfo{author}{\bibfnamefont{U.}~\bibnamefont{Bauer}},
  \bibinfo{author}{\bibfnamefont{S.}~\bibnamefont{Emori}}, \bibnamefont{and}
  \bibinfo{author}{\bibfnamefont{G.~S.~D.} \bibnamefont{Beach}},
  \bibinfo{journal}{Nat. Nanotechnol.} \textbf{\bibinfo{volume}{8}},
  \bibinfo{pages}{411} (\bibinfo{year}{2013}).

\bibitem[{\citenamefont{Bauer et~al.}(2015)\citenamefont{Bauer, Yao, Tan,
  Agrawal, Emori, Tuller, van Dijken, and Beach}}]{Bauer2015}
\bibinfo{author}{\bibfnamefont{U.}~\bibnamefont{Bauer}},
  \bibinfo{author}{\bibfnamefont{L.}~\bibnamefont{Yao}},
  \bibinfo{author}{\bibfnamefont{A.~J.} \bibnamefont{Tan}},
  \bibinfo{author}{\bibfnamefont{P.}~\bibnamefont{Agrawal}},
  \bibinfo{author}{\bibfnamefont{S.}~\bibnamefont{Emori}},
  \bibinfo{author}{\bibfnamefont{H.~L.} \bibnamefont{Tuller}},
  \bibinfo{author}{\bibfnamefont{S.}~\bibnamefont{van Dijken}},
  \bibnamefont{and} \bibinfo{author}{\bibfnamefont{G.~S.~D.}
  \bibnamefont{Beach}}, \bibinfo{journal}{Nat. Mater.}
  \textbf{\bibinfo{volume}{14}}, \bibinfo{pages}{174} (\bibinfo{year}{2015}).

\bibitem[{\citenamefont{Chu et~al.}(2008)\citenamefont{Chu, Martin, Holcomb,
  Gajek, Han, He, Balke, Yang, Lee, Hu et~al.}}]{Chu}
\bibinfo{author}{\bibfnamefont{Y.-H.} \bibnamefont{Chu}},
  \bibinfo{author}{\bibfnamefont{L.~W.} \bibnamefont{Martin}},
  \bibinfo{author}{\bibfnamefont{M.~B.} \bibnamefont{Holcomb}},
  \bibinfo{author}{\bibfnamefont{M.}~\bibnamefont{Gajek}},
  \bibinfo{author}{\bibfnamefont{S.-J.} \bibnamefont{Han}},
  \bibinfo{author}{\bibfnamefont{Q.}~\bibnamefont{He}},
  \bibinfo{author}{\bibfnamefont{N.}~\bibnamefont{Balke}},
  \bibinfo{author}{\bibfnamefont{C.-H.} \bibnamefont{Yang}},
  \bibinfo{author}{\bibfnamefont{D.}~\bibnamefont{Lee}},
  \bibinfo{author}{\bibfnamefont{W.}~\bibnamefont{Hu}}, \bibnamefont{et~al.},
  \bibinfo{journal}{Nature Mater.} \textbf{\bibinfo{volume}{7}},
  \bibinfo{pages}{478} (\bibinfo{year}{2008}).

\bibitem[{\citenamefont{Lebeugle et~al.}(2009)\citenamefont{Lebeugle, Mougin,
  Viret, Colson, and Ranno}}]{Lebeugle}
\bibinfo{author}{\bibfnamefont{D.}~\bibnamefont{Lebeugle}},
  \bibinfo{author}{\bibfnamefont{A.}~\bibnamefont{Mougin}},
  \bibinfo{author}{\bibfnamefont{M.}~\bibnamefont{Viret}},
  \bibinfo{author}{\bibfnamefont{D.}~\bibnamefont{Colson}}, \bibnamefont{and}
  \bibinfo{author}{\bibfnamefont{L.}~\bibnamefont{Ranno}},
  \bibinfo{journal}{Phys. Rev. Lett.} \textbf{\bibinfo{volume}{103}},
  \bibinfo{pages}{257601} (\bibinfo{year}{2009}).

\bibitem[{\citenamefont{Heron et~al.}(2011)\citenamefont{Heron, Trassin,
  Ashraf, Gajek, He, Yang, Nikonov, Chu, Salahuddin, and Ramesh}}]{Heron}
\bibinfo{author}{\bibfnamefont{J.~T.} \bibnamefont{Heron}},
  \bibinfo{author}{\bibfnamefont{M.}~\bibnamefont{Trassin}},
  \bibinfo{author}{\bibfnamefont{K.}~\bibnamefont{Ashraf}},
  \bibinfo{author}{\bibfnamefont{M.}~\bibnamefont{Gajek}},
  \bibinfo{author}{\bibfnamefont{Q.}~\bibnamefont{He}},
  \bibinfo{author}{\bibfnamefont{S.}~\bibnamefont{Yang}},
  \bibinfo{author}{\bibfnamefont{D.~E.} \bibnamefont{Nikonov}},
  \bibinfo{author}{\bibfnamefont{Y.-H.} \bibnamefont{Chu}},
  \bibinfo{author}{\bibfnamefont{S.}~\bibnamefont{Salahuddin}},
  \bibnamefont{and} \bibinfo{author}{\bibfnamefont{R.}~\bibnamefont{Ramesh}},
  \bibinfo{journal}{Phys. Rev. Lett.} \textbf{\bibinfo{volume}{107}},
  \bibinfo{pages}{217202} (\bibinfo{year}{2011}).

\bibitem[{\citenamefont{You et~al.}(2013)\citenamefont{You, Wang, Zou, Lim,
  Zhou, Ding, Chen, and Wang}}]{You}
\bibinfo{author}{\bibfnamefont{L.}~\bibnamefont{You}},
  \bibinfo{author}{\bibfnamefont{B.}~\bibnamefont{Wang}},
  \bibinfo{author}{\bibfnamefont{X.}~\bibnamefont{Zou}},
  \bibinfo{author}{\bibfnamefont{Z.~S.} \bibnamefont{Lim}},
  \bibinfo{author}{\bibfnamefont{Y.}~\bibnamefont{Zhou}},
  \bibinfo{author}{\bibfnamefont{H.}~\bibnamefont{Ding}},
  \bibinfo{author}{\bibfnamefont{L.}~\bibnamefont{Chen}}, \bibnamefont{and}
  \bibinfo{author}{\bibfnamefont{J.}~\bibnamefont{Wang}},
  \bibinfo{journal}{Phys. Rev. B} \textbf{\bibinfo{volume}{88}},
  \bibinfo{pages}{184426} (\bibinfo{year}{2013}).

\bibitem[{\citenamefont{Lahtinen et~al.}(2011)\citenamefont{Lahtinen, Tuomi,
  and van Dijken}}]{Lahtinen2011}
\bibinfo{author}{\bibfnamefont{T.~H.~E.} \bibnamefont{Lahtinen}},
  \bibinfo{author}{\bibfnamefont{J.~O.} \bibnamefont{Tuomi}}, \bibnamefont{and}
  \bibinfo{author}{\bibfnamefont{S.}~\bibnamefont{van Dijken}},
  \bibinfo{journal}{Adv. Mater.} \textbf{\bibinfo{volume}{23}},
  \bibinfo{pages}{3187} (\bibinfo{year}{2011}).

\bibitem[{\citenamefont{Lahtinen et~al.}(2012)\citenamefont{Lahtinen, Franke,
  and van Dijken}}]{Lahtinen2012}
\bibinfo{author}{\bibfnamefont{T.~H.~E.} \bibnamefont{Lahtinen}},
  \bibinfo{author}{\bibfnamefont{K.~J.~A.} \bibnamefont{Franke}},
  \bibnamefont{and} \bibinfo{author}{\bibfnamefont{S.}~\bibnamefont{van
  Dijken}}, \bibinfo{journal}{Sci. Rep.} \textbf{\bibinfo{volume}{2}},
  \bibinfo{pages}{258} (\bibinfo{year}{2012}).

\bibitem[{\citenamefont{Franke et~al.}(2012)\citenamefont{Franke, Lahtinen, and
  van Dijken}}]{Franke2012}
\bibinfo{author}{\bibfnamefont{K.~J.~A.} \bibnamefont{Franke}},
  \bibinfo{author}{\bibfnamefont{T.~H.~E.} \bibnamefont{Lahtinen}},
  \bibnamefont{and} \bibinfo{author}{\bibfnamefont{S.}~\bibnamefont{van
  Dijken}}, \bibinfo{journal}{Phys. Rev. B} \textbf{\bibinfo{volume}{85}},
  \bibinfo{pages}{094423} (\bibinfo{year}{2012}).

\bibitem[{\citenamefont{Chopdekar et~al.}(2012)\citenamefont{Chopdekar, Malik,
  Rodr\'{i}guez, Guyader, Takamura, Scholl, Stender, Schneider, Bernhard,
  Nolting et~al.}}]{Chopdekar}
\bibinfo{author}{\bibfnamefont{R.~V.} \bibnamefont{Chopdekar}},
  \bibinfo{author}{\bibfnamefont{V.~K.} \bibnamefont{Malik}},
  \bibinfo{author}{\bibfnamefont{A.~F.} \bibnamefont{Rodr\'{i}guez}},
  \bibinfo{author}{\bibfnamefont{L.~L.} \bibnamefont{Guyader}},
  \bibinfo{author}{\bibfnamefont{Y.}~\bibnamefont{Takamura}},
  \bibinfo{author}{\bibfnamefont{A.}~\bibnamefont{Scholl}},
  \bibinfo{author}{\bibfnamefont{D.}~\bibnamefont{Stender}},
  \bibinfo{author}{\bibfnamefont{C.~W.} \bibnamefont{Schneider}},
  \bibinfo{author}{\bibfnamefont{C.}~\bibnamefont{Bernhard}},
  \bibinfo{author}{\bibfnamefont{F.}~\bibnamefont{Nolting}},
  \bibnamefont{et~al.}, \bibinfo{journal}{Phys. Rev B}
  \textbf{\bibinfo{volume}{86}}, \bibinfo{pages}{014408}
  (\bibinfo{year}{2012}).

\bibitem[{\citenamefont{Streubel et~al.}(2013)\citenamefont{Streubel,
  K\"{o}hler, Sch\"{a}fer, and Eng}}]{Streubel}
\bibinfo{author}{\bibfnamefont{R.}~\bibnamefont{Streubel}},
  \bibinfo{author}{\bibfnamefont{D.}~\bibnamefont{K\"{o}hler}},
  \bibinfo{author}{\bibfnamefont{R.}~\bibnamefont{Sch\"{a}fer}},
  \bibnamefont{and} \bibinfo{author}{\bibfnamefont{L.~M.} \bibnamefont{Eng}},
  \bibinfo{journal}{Phys. Rev. B} \textbf{\bibinfo{volume}{87}},
  \bibinfo{pages}{054410} (\bibinfo{year}{2013}).

\bibitem[{\citenamefont{Franke et~al.}(2014)\citenamefont{Franke, {L\'{o}pez
  Gonz\'{a}lez}, H\"{a}m\"{a}l\"{a}inen, and van Dijken}}]{Franke2014}
\bibinfo{author}{\bibfnamefont{K.~J.~A.} \bibnamefont{Franke}},
  \bibinfo{author}{\bibfnamefont{D.}~\bibnamefont{{L\'{o}pez Gonz\'{a}lez}}},
  \bibinfo{author}{\bibfnamefont{S.~J.} \bibnamefont{H\"{a}m\"{a}l\"{a}inen}},
  \bibnamefont{and} \bibinfo{author}{\bibfnamefont{S.}~\bibnamefont{van
  Dijken}}, \bibinfo{journal}{Phys. Rev. Lett.} \textbf{\bibinfo{volume}{112}},
  \bibinfo{pages}{017201} (\bibinfo{year}{2014}).

\bibitem[{\citenamefont{Merz}(1954)}]{Merz}
\bibinfo{author}{\bibfnamefont{W.~J.} \bibnamefont{Merz}},
  \bibinfo{journal}{Phys. Rev.} \textbf{\bibinfo{volume}{95}},
  \bibinfo{pages}{690} (\bibinfo{year}{1954}).

\bibitem[{\citenamefont{Hubert}(1979)}]{Hubert1979}
\bibinfo{author}{\bibfnamefont{A.}~\bibnamefont{Hubert}},
  \bibinfo{journal}{IEEE Trans. Magn.} \textbf{\bibinfo{volume}{15}},
  \bibinfo{pages}{1251} (\bibinfo{year}{1979}).

\bibitem[{\citenamefont{Hubert and Sch\"{a}fer}(1998)}]{Hubert1998}
\bibinfo{author}{\bibfnamefont{A.}~\bibnamefont{Hubert}} \bibnamefont{and}
  \bibinfo{author}{\bibfnamefont{R.}~\bibnamefont{Sch\"{a}fer}},
  \emph{\bibinfo{title}{Magnetic Domains. The Analysis of Magnetic
  Microstructures}} (\bibinfo{publisher}{Springer-Verlag, Berlin Heidelberg},
  \bibinfo{year}{1998}).

\bibitem[{\citenamefont{Favieres et~al.}(2013)\citenamefont{Favieres, Vergara,
  and Madurga}}]{Favieres}
\bibinfo{author}{\bibfnamefont{C.}~\bibnamefont{Favieres}},
  \bibinfo{author}{\bibfnamefont{J.}~\bibnamefont{Vergara}}, \bibnamefont{and}
  \bibinfo{author}{\bibfnamefont{V.}~\bibnamefont{Madurga}},
  \bibinfo{journal}{J. Phys.: Condens. Matter} \textbf{\bibinfo{volume}{25}},
  \bibinfo{pages}{066002} (\bibinfo{year}{2013}).

\bibitem[{\citenamefont{Berger and Oepen}(1992)}]{Berger}
\bibinfo{author}{\bibfnamefont{A.}~\bibnamefont{Berger}} \bibnamefont{and}
  \bibinfo{author}{\bibfnamefont{H.~P.} \bibnamefont{Oepen}},
  \bibinfo{journal}{Phys. Rev. B} \textbf{\bibinfo{volume}{45}},
  \bibinfo{pages}{12596(R)} (\bibinfo{year}{1992}).

\bibitem[{\citenamefont{Fr\"{o}mter et~al.}(2011)\citenamefont{Fr\"{o}mter,
  Hankemeier, Oepen, and Kirschner}}]{Fromter}
\bibinfo{author}{\bibfnamefont{R.}~\bibnamefont{Fr\"{o}mter}},
  \bibinfo{author}{\bibfnamefont{S.}~\bibnamefont{Hankemeier}},
  \bibinfo{author}{\bibfnamefont{H.~P.} \bibnamefont{Oepen}}, \bibnamefont{and}
  \bibinfo{author}{\bibfnamefont{J.}~\bibnamefont{Kirschner}},
  \bibinfo{journal}{Rev. Sci. Instrum.} \textbf{\bibinfo{volume}{82}},
  \bibinfo{pages}{033704} (\bibinfo{year}{2011}).

\bibitem[{\citenamefont{Lilley}(1950)}]{Lilley}
\bibinfo{author}{\bibfnamefont{B.}~\bibnamefont{Lilley}},
  \bibinfo{journal}{Philosophical Magazine and Journal of Science: Series 7}
  \textbf{\bibinfo{volume}{41}}, \bibinfo{pages}{792} (\bibinfo{year}{1950}).

\bibitem[{\citenamefont{Johnson et~al.}(1996)\citenamefont{Johnson, Bloemen,
  den Broeder, and de~Vries}}]{Johnson}
\bibinfo{author}{\bibfnamefont{M.~T.} \bibnamefont{Johnson}},
  \bibinfo{author}{\bibfnamefont{P.~J.~H.} \bibnamefont{Bloemen}},
  \bibinfo{author}{\bibfnamefont{F.~J.~A.} \bibnamefont{den Broeder}},
  \bibnamefont{and} \bibinfo{author}{\bibfnamefont{J.~J.}
  \bibnamefont{de~Vries}}, \bibinfo{journal}{Rep. Prog. Phys.}
  \textbf{\bibinfo{volume}{59}}, \bibinfo{pages}{1409} (\bibinfo{year}{1996}).

\bibitem[{\citenamefont{Nogu\'{e}s and Schuller}(1999)}]{Nogues}
\bibinfo{author}{\bibfnamefont{J.}~\bibnamefont{Nogu\'{e}s}} \bibnamefont{and}
  \bibinfo{author}{\bibfnamefont{I.~K.} \bibnamefont{Schuller}},
  \bibinfo{journal}{J. Magn. Magn. Mater.} \textbf{\bibinfo{volume}{192}},
  \bibinfo{pages}{203} (\bibinfo{year}{1999}).

\end{thebibliography}

\end{document}